# Probing molecular dynamics at the nanoscale via an individual paramagnetic center

T. Staudacher[1,2], N. Raatz[3], S. Pezzagna[3], J. Meijer[3], F. Reinhard[1, 4], C. A. Meriles[5,*], and J. Wrachtrup[1]

[1]3rd physics Institute and IQST, University of Stuttgart, 70569 Stuttgart, Germany

[2]Max Planck Institute for Solid State Research, 70174 Stuttgart, Germany

[3]Department of Nuclear Solid State Physics, Institute for Experimental Physics II, University of Leipzig, Linnéstr. 5, 04103 Leipzig, Germany

[4]TUM – Technical University of Munich, Walter Schottky Institut, Am Coulombwall 4, 85748 Garching, Germany

[5]Department of Physics, CUNY-City College of New York, 160 Convent Ave, New York, NY 10031, USA

*For correspondence: cmeriles@sci.ccny.cuny.edu

## Abstract

**Understanding the dynamics of molecules adsorbed to surfaces or confined to small volumes is a matter of increasing scientific and technological importance. Here, we demonstrate a pulse protocol using individual paramagnetic nitrogen vacancy (NV) centers in diamond to observe the time evolution of $^1H$ spins from organic molecules located a few nanometers from the diamond surface. The protocol records temporal correlations among the interacting $^1H$ spins, and thus is sensitive to the local system dynamics via its impact on the nuclear spin relaxation and interaction with the NV. We are able to gather information on the nanoscale rotational and translational diffusion dynamics by carefully analyzing the time dependence of the NMR signal. Applying this technique to various liquid and solid samples, we find evidence that liquid samples form a semi-solid layer of 1.5 nm thickness on the surface of diamond, where translational diffusion is suppressed while rotational diffusion remains present. Extensions of the present technique could be adapted to highlight the chemical composition of molecules tethered to the diamond surface or to investigate thermally or chemically activated dynamical processes such as molecular folding.**

*Main text*

Nuclear magnetic resonance (NMR) is among the most versatile tools for investigating the dynamics of molecular processes down to the atomic level. It is widely used in physical and life sciences, but has been limited to large sample quantities due to the low sensitivity of conventional detection methods[1]. Performing NMR detection at the nanoscale can substantially expand the microscopist's toolbox, potentially allowing for non-destructive imaging of complex macromolecules and/or studying the dynamics of diverse biochemical systems. One route to performing nanoNMR is magnetic resonance force microscopy (MRFM), already used to image small organisms with nanometer resolution[2]. Typical operating conditions of MRFM however require ultralow temperatures and high vacuum[2, 3], which, unfortunately, are incompatible with most molecular processes of interest.

An alternate approach to NMR at the nanoscale makes use of individually addressable paramagnetic centers near the surface of a solid-state host to probe sample spin species in its immediate vicinity. Perhaps the most prominent example is the nitrogen-vacancy (NV) center, a spin-1 defect in the diamond lattice formed by a substitutional nitrogen atom and an adjacent vacancy[4]. Recently, single NV centers separated only a few nanometers from the diamond surface were used to detect the NMR signal associated with the random magnetic spin noise of a nanoscale proton ensemble under ambient conditions[5, 6]. Subsequent studies extended this initial work to other spin species and demonstrated improved detection sensitivity, attaining the limit of a few nuclear spins[7, 8, 9, 10, 11]. Furthermore, by articulating NV magnetometry with scanning microscopy, it has been possible to image the spatial distribution of nuclear spins within a polymeric phantom with about 10 nm resolution[12, 13].

In this study, we use shallow NVs to probe mesoscale proton ensembles from different organic substances deposited on the diamond surface. We resort to a form of correlation spectroscopy and reconstruct the equivalent of a nuclear 'free-induction-decay' (FID), which, unlike the NMR counterpart, does not require nuclear spin pre-polarization. This pseudo FID—below referred to as 'correlation signal'—has a limited decay time governed by the NV spin-lattice relaxation time $T_1$ (typically longer than the NV coherence lifetime $T_2$), which allows us to attain spectral resolution superior to that possible with standard magnetometry techniques. Upon applying this scheme to solid- and liquid-state substances we find substantial differences in the correlation signal envelope, which we associate with the presumably dissimilar molecular dynamics governing these systems. In particular, we observe long-lived $^1$H signals from oil molecules, which we interpret in terms of an interplay between molecular tumbling and self-diffusion.

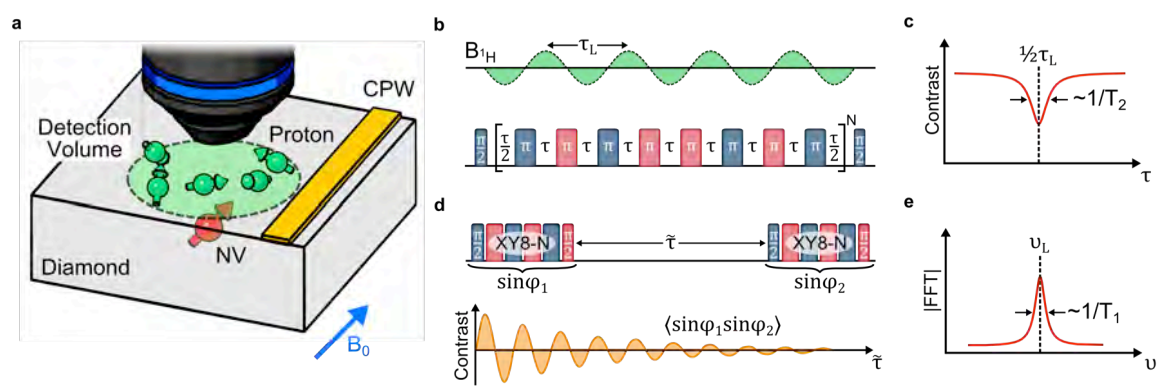

**Figure 1. Schematics of the experimental setup and basic detection protocol.** (a) An organic sample is brought into contact with the diamond surface, and shallow NV centers are used as NMR detectors. (b) XY8-N multi-pulse sequence. A change in the NV response is observed when the inter-pulse separation $\tau$ matches half the Larmor period. The blue/red hue indicates different MW phases, which are shifted 90° relative to each other. (c) Repeating the XY8-N sequence for multiple pulse spacings $\tau$ yields an effective CW spectrum of the nuclear spin noise. (d) Nuclear spin correlation protocol. Selective detection of the proton spins is attained by choosing the pulse spacing $\tau$ equal to half the Larmor period. The correlation signal shows a decaying oscillating behavior as a function of the $^1$H spin evolution interval $\tilde{\tau}$. (e) The Fourier transform of the time domain signal yields the sample NMR spectrum.

Fig. 1a depicts a typical NMR experiment using NV centers. The setup consists of a home-built confocal microscope, which excites single NV centers in the illumination volume via a 532 nm laser. Green light initializes the NV—a spin-1 system—into the $m_s$=0 level of its ground state triplet, which features a zero-field splitting of 2.87 GHz[14]. The spin-state dependent back fluorescence of the NV center is collected via the imaging objective and is focused onto a single photon detector. A moveable electromagnet provides a static magnetic field of about ~25 mT and a nearby coplanar waveguide (CPW) is used for the application of microwave irradiation at the frequency of the $m_S = 0 \leftrightarrow m_S = -1$ transition (~2.2 GHz). For the experiments herein we use single NV centers produced via 2.5 keV $^{15}N^+$ ion implantation into a type-IIa [100] diamond crystal (Section A of the Supplementary Information). The dipolar coupling to nuclear spins external to the diamond lattice is strong enough to imprint the NV response with a signature that originates from the statistical nuclear spin polarization[5, 6]. The detection volume is roughly defined by the NV distance to the surface — about 5 nm in the present case — which approximately corresponds to ~$10^3$ protons for typical organic samples[5, 6, 10].

A common way to detect the NMR signal via the NV center is based on a quantum lock-in algorithm[15, 16], which is implemented through an XY8-N dynamical decoupling sequence[6 - 12]. Here a train of equidistant $\pi$-pulses is used to selectively enhance the NV detection sensitivity at a frequency determined by the inverse pulse spacing $\tau^{-1}$. As shown in Fig. 1b, the XY8-N sequence is embedded within a Ramsey protocol (comprising two $\pi/2$-pulses), so as to convert the integrated effect of the nuclear spins — in the form of an accumulated NV phase shift — into a change in the NV fluorescence. The result of such a measurement is an effective CW spectrum of the sample spins (Fig. 1c), whose linewidth is ultimately limited by the XY8-N coherence time $T_2^{(NV)}$.

The coherence lifetimes of shallow NVs are often times shorter than the characteristic time scales governing nuclear spins, thus complicating our ability to gather detailed spectroscopic information on the structure, chemical composition, or dynamics of the system under investigation. One way to circumvent this limitation is the nuclear spin detection protocol of Fig. 1d[17] designed to exploit the typically longer NV spin-lattice relaxation times $T_1^{(NV)}$. The pulse sequence comprises two XY8-N trains separated by a variable interval $\tilde{\tau}$; in each of them the interpulse separation is kept in sync with the sample spin Larmor precession, i.e., we choose $\tau = \frac{1}{2}\tau_L$. During the evolution time $\tilde{\tau}$ the magnetization information is stored in the longitudinal NV spin component while the sample nuclear spins are allowed to evolve. The underlying idea is that if the nuclear spin coherence loss is sufficiently slow, the phases $\phi_1,\ \phi_2$ picked up by the NV during the two consecutive XY8-N interrogations are correlated with each other. The latter results in a signal similar to an FID in conventional NMR, which, however, does not require nuclear spin (pre-)polarization. Interestingly, nuclear spin coherences lasting up to hundreds of microseconds — the typical $T_1^{(NV)}$ time of the NVs we use here  (Section F of the Supplementary Information)— can be probed with this technique, thus allowing us to better discriminate between different sample dynamics. In this light, our technique can be considered an alternative to double resonance schemes already used to reconstruct FID-like signals from [^1]H spins near shallow NVs[5].

Fig. 2a shows the NV response in the presence of a solid organic film as a function of the evolution time $\tilde{\tau}$. This system — a complex mixture of long-chain polymers hereby referred to as Sample A — is formed by the adhesive used to affix the diamond crystal to the sample holder  (Merckoglas®). Similar to a conventional FID the correlation signal oscillates over tens of microseconds to gradually decay to zero. We find that the decay is reasonably described by an exponential, and has a characteristic time constant $T_{corr}^{(A)} \sim 20$ μs. After a cosine transformation, we find a peak centered at the [^1]H Larmor frequency (~1.06 MHz) exhibiting a Lorentzian linewidth of ~30 kHz, in agreement with that expected for static protons in a typical solid-state organic system[18]. Of note, the correlation signal amplitude is only a small fraction (~10 %) of the maximum possible fluorescence contrast (30 % between spin states $m_S = 0$ and $m_S = \pm 1$)[14]. On the other hand, the resulting spectral linewidth is about a factor 2 smaller than that obtained with an XY8-10 sequence, which highlights the limitations inherent to sensing protocols governed by the coherence lifetime of the probe NV (Fig. 2b).

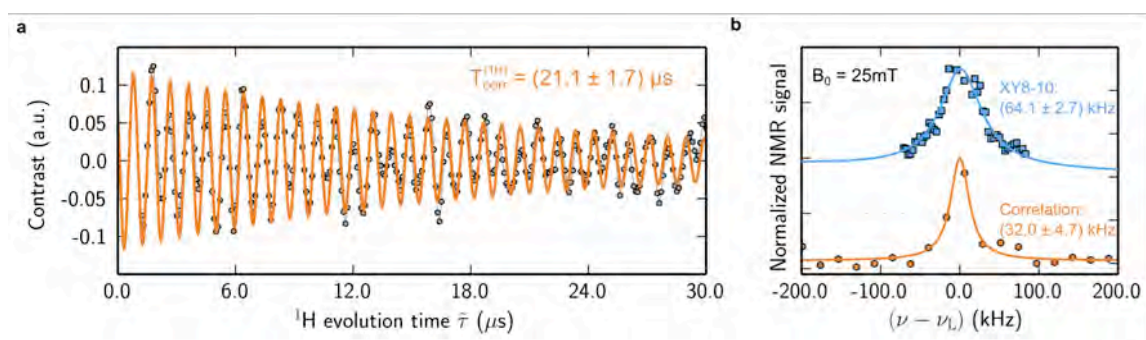

**Figure 2. Time-resolved NMR of near-surface [^1]H spins.** (a) XY8-3 correlation signal (circles) from protons in a solid polymeric mixture (Sample A) in contact with the diamond surface. The yellow trace corresponds to an exponentially-damped sinusoid with time constant of 21 μs and frequency equal to ~1 MHz. The vertical axis indicates the contrast in a scale relative to the NV Rabi amplitude. (b) Comparison between the [^1]H-NMR spectra obtained with an XY8-10 sequence (blue squares) and after Fourier transformation of the XY8-3 correlation protocol shown in (a) (yellow circles). The blue and yellow traces respectively indicate Lorentzian fits to each data set, accompanied by their corresponding FWHM values. Both curves are vertically displaced for clarity.

Since the time scale probed in Fig. 2a is still considerably shorter than $T_1^{(NV)}$ (Section F of the Supplementary Information), a natural question is whether this technique can be exploited to investigate longer-lived nuclear spin coherences arising, for example, from alternate forms of motional narrowing. A first step in this direction is shown in Fig. 3 where we compare representative correlation signals from different organic systems, including that in Fig. 2 as well as a softer polymeric film (Polydimethylsiloxane, a.k.a PDMS) and a drop of immersion oil (Fluka Analytical, 10976) in direct contact with the diamond surface (Sections A and B of the Supplementary Information). Below we refer to the latter two systems as Samples B and C, respectively. In all three cases we observe periodic oscillations reflecting the already highlighted precession of proton spins about the applied magnetic field. However, the time over which these oscillations last and, perhaps more importantly, the way the signal envelope changes over time differ significantly in each case. In particular we find that the correlation signal from Sample C exhibits a long-lasting tail extending to about 80 μs (possibly limited by NV spin lattice relaxation, Section F of the Supplementary Material). Unlike Sample A, this behavior cannot be captured by a single-exponential envelope (see the lower insert in Fig. 3) and thus points to differing nuclear spin relaxation mechanisms. Sample B, on the other hand, shows a somewhat intermediate, longer-lived signal, which, nonetheless, does not depart from the single-exponential response observed in Sample A.

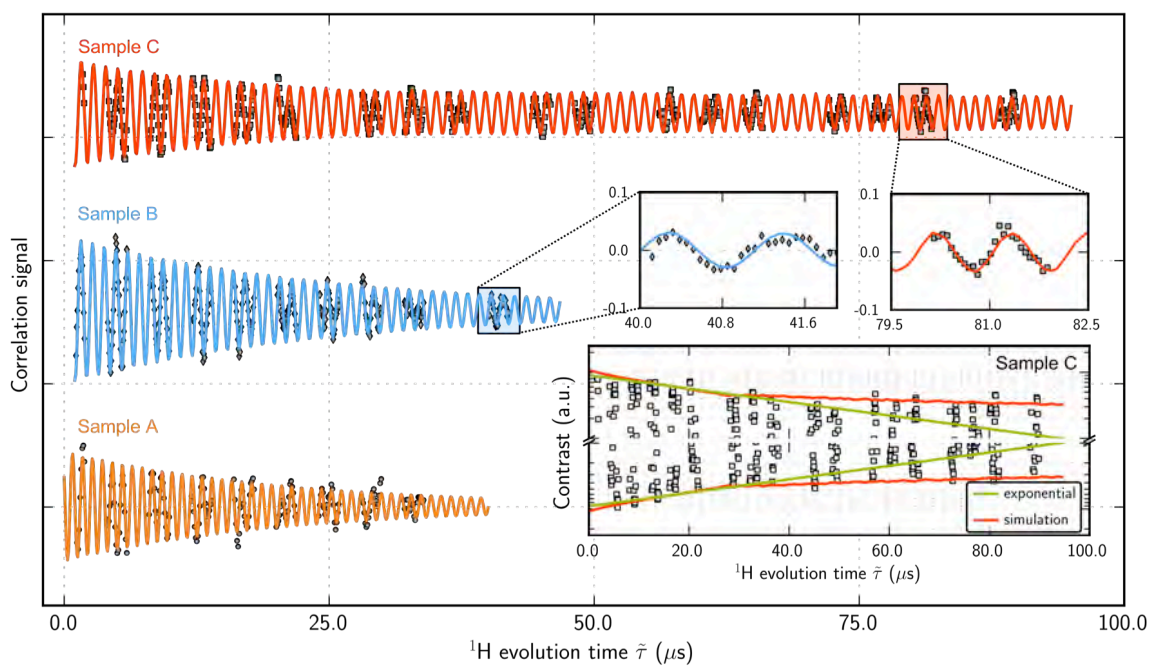

**Figure 3. XY8-3 correlation signals for three different organic samples.** The data corresponding to both solid samples (Samples A and B, respectively orange and blue traces) match an exponentially-damped sinusoid with time constants of ~21 µs and ~29 µs. By contrast, the correlation signal from protons in Sample C exhibits a long-lived tail that outlives the exponential decay at early times (green trace in the figure inset). The overall response can be reproduced semi-quantitatively via a model comprising a 1.5 nm layer of adsorbed molecules rotating about fixed positions and an outer section of self-diffusing fluid. Best agreement with the experimental observations is attained assuming translational and rotational diffusion constants of 0.3 nm²/µs and 0.05 rad²/µs for the outer and inner layer, respectively (red sinusoid trace in the main figure, and red envelope in the figure insert). Given the relatively low fluorescence contrast, only portions of the signal were measured. Though obtained with different NVs, each curve must be considered representative of the NV response for the corresponding proton ensemble under study (Section B of the SI).

To gain a more quantitative understanding of the mechanisms at play, we start by modeling the NV response in a way that accommodates the different dynamics governing solid and liquid samples. Using a semi-classical approximation to describe the NV interaction with the nuclear spin bath (Section C of the Supplementary Information) and assuming all molecules move independently, we write the correlation signal as

$$S(\tau, \tilde{\tau}) \sim \sum_{i,j} \langle \phi_{i,j}(0), \phi_{i,j}(\tilde{\tau}) \rangle \sim \cos\left(2\pi \nu_L (2N_{\text{tot}}\tau + \tilde{\tau})\right) \sum_{i,j} A_{i,j}(\tau, \tilde{\tau}) \, e^{-\tilde{\tau}/T_{corr}^{(i,j)}}, \qquad (1)$$

where the sum extends over all proton spins $i$ in the $j$-th molecule and $\phi_{i,j}$ denotes the corresponding contribution to the accumulated NV phase, $A_{i,j}$ is the resulting signal amplitude, $\nu_L$ is the nuclear Larmor frequency, and $N_{\text{tot}}$ is the total number of $\pi$-pulses in each XY8-N train. The signal amplitude $A_{i,j}$ is proportional to the rms-field generated by the proton spins at the positon of the NV center and varies with different molecule positions and/or orientations over the evolution time $\tilde{\tau}$. In the case of a solid, molecules occupy fixed positions and nuclear spin relaxation is dominated by the nuclear dipolar interactions, presumably homogeneous throughout the sample. In this limit, the rate $1/T_{corr}^{(i,j)}$ describing the spin relaxation of nuclear moment $\mu_{i,j}$ approaches a uniform value, and Eq. (1) converges to an exponentially damped sinusoid[17], the case observed for both solid samples. For Sample A, we find $T_{corr}^{(A)} \sim 21$ μs which corresponds to a decay dominated by static dipolar couplings, of order $\mu^2/|\Delta r|^3 \sim 30$ kHz for typical inter-proton distances $|\Delta r| \sim 0.1$ nm in organic samples. The slightly longer coherence lifetime in Sample B, $T_{corr}^{(B)} \sim 29$ μs, possibly originates from an enhanced mobility of the molecular mobility in this material (see below, and Section G of the Supplementary Information).

Fluidic systems, on the other hand, differ from solids in that molecules experience a markedly distinct dynamics dominated by fast tumbling and self-diffusion. Both mechanisms contribute to average out the inter-nuclear spin couplings and thus lead to longer nuclear spin coherence lifetimes. However, given the nanoscale detection volume of a shallow NV — roughly restricted to a half-sphere of diameter equal to the NV depth[6,19] — molecules interacting with the probe paramagnetic center may exchange with the bulk of the system at some arbitrary time during the protocol. This situation is somewhat reminiscent of that found in fluorescence correlation spectroscopy (FCS), where molecules diffuse in and out of the detection volume[20, 21, 22]. In the present case, we simplify the problem by assuming that molecules occupy 'frozen' positions during each interrogation interval, thus 'instantaneously' tagging the NV with a phase shift corresponding to the molecule positions at times 0 and $\tilde{\tau}$. In this limit, we rewrite the NV response as

$$S(\tau, \tilde{\tau}) \sim \cos\left(2\pi\nu_L(2N_{tot}\tau + \tilde{\tau})\right) \sum_{i,j} A_{i,j}^{(0)} \, p_{i,j}(\tilde{\tau}) \, e^{-\tau/T_2^{(NV)}} e^{-\tilde{\tau}/T_2^{(i,j)}}, \qquad (2)$$

where $A_{i,j}^{(0)} \equiv A_{i,j}(0,0)$, $p_{i,j}(\tilde{\tau})$ denotes the conditional probability of nuclear spin $(i, j)$ remaining within the detection volume over the correlation interval $\tilde{\tau}$, and $T_2^{(i,j)}$ is the nuclear spin transverse relaxation time. The correlation decay of Eq. (1) $e^{-\tilde{\tau}/T_{corr}^{(i,j)}}$ is expressed as the product of the NV decay $e^{-\tau/T_2^{(NV)}}$, the conditional probability $p_{i,j}(\tilde{\tau})$, and the NMR decay of the sample spins $e^{-\tilde{\tau}/T_2^{(i,j)}}$. For simplicity, we assumed in Eq. (2) that the signal amplitude $A_{i,j}$, and $T_2^{(i,j)}$ remain unchanged after self diffusion during $\tilde{\tau}$, which is consistent with a bimodal distribution where molecules either self-diffuse or not depending on their location at time zero.

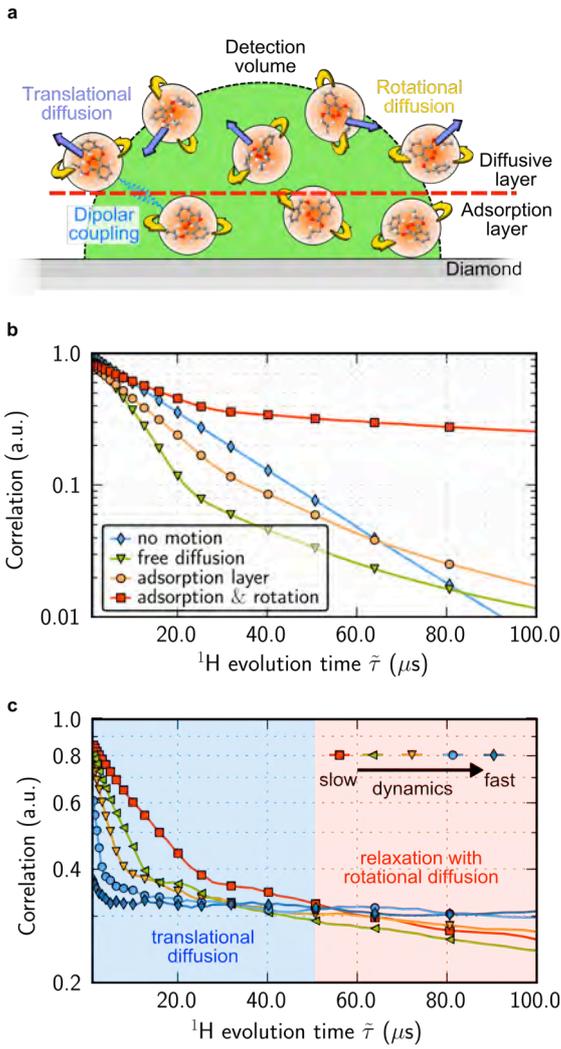

Although the dynamics of molecules at a solid-liquid interface are not fully understood, several studies suggest that solids induce order in adjacent fluids[23]. The boundary condition typically invoked is one where the liquid is (nearly) static over the surface[24]. In particular, recent AFM experiments in water indicate that the transition to bulk fluid dynamics is abrupt and takes place on the nm range, depending on the surface hydrophobicity[25]. To numerically calculate $S(\tau, \tilde{\tau})$ in Eq. (2) we divide the detection volume in two layers (Fig. 4a): Molecules adjacent to the surface rotate about fixed positions, while molecules in the outer layer rotate and physically diffuse away from the NV. The set of conditional probabilities $p_{i,j}(\tilde{\tau})$ for the mobile layer is calculated from a Monte Carlo run assuming a bulk diffusion coefficient for the fluid. To bring the number of free parameters to a minimum, we tie the rotation rate of bulk molecules to the diffusion constant via the Stokes-Einstein and Debye-Stokes-Einstein equations; comparison between the observed and calculated signals is carried out by systematically varying the rigid layer thickness and self-diffusion constant of the fluid (Sections D and E in the Supplementary Material)

To illustrate the impact of the dynamics of molecular diffusion and rotation on the correlation signal, we use Eq. (2) to calculate the envelope governing the decay of $S(\tau, \tilde{\tau})$ assuming different boundary conditions. For example, the green trace in Fig. 4b shows the response anticipated in the case where the sample dynamics corresponds to that of a bulk fluid up to the very surface of the diamond crystal, i.e., we assume the layer of adsorbed molecules has negligible thickness. Using a diffusion constant $D_{\text{oil}} \sim 0.3 \times 10^{-12}$ m$^2$/s (comparable to that anticipated for this sample) we find that the calculated signal envelope somewhat reproduces the trend in our experiment, namely, it exhibits a long lasting tail that outlives the exponential decay at early times (green trace in Fig. 4b). A numerical analysis shows that this tail stems from the longer nuclear spin coherences inherent to the mobile molecules of a fluid. Quantitatively, however, the calculated envelope does not agree well with the experimental data: First we note that the relative contribution of the long-lasting signal is comparatively small (because most molecules leave the detection volume when $\tilde{\tau}$ is sufficiently long). Further, compared to the case where no motion is present (infinite rigid layer, blue trace in Fig. 4b), we find a much faster initial decay (also absent in the experimental signal of Fig. 3). Changing the diffusion constant to greater or smaller values either accelerates the initial decay or further reduces the tail, thus pointing to the inadequacy of the underlying model (Section E of the Supplementary Information).

We can, however, reproduce the experimental data when we assume the presence of a 1.5 nm thick adsorption layer (orange trace in Fig. 4b), particularly if molecules within this layer are allowed to rotate about their equilibrium positions (red trace in Fig. 4b). The observed and calculated correlation signals for the fluidic sample are superimposed in Fig. 3b (red trace). The result captures reasonably well both the initial decay rate and the amplitude of the long-lived contribution. From the transverse relaxation rate $1/T_{corr}^{(long)} \approx 3$ kHz assigned to adsorbed protons we determine the rotational correlation time $\tau_R^{(Ads)} \approx 3.3$ μs, much longer than in the outer segment of the sample ($\tau_R^{(Out)} \sim 190$ ns as calculated from the Stokes-Einstein equation for $D_{oil} \sim 0.3 \times 10^{-12}$ m²/s, Section C of the Supplementary Information). We caution, however, that this value must be interpreted as an upper limit given the likely impact of NV spin-lattice relaxation on the correlation signal of Sample C (Section F of the Supplementary Information).

In light of the above discussion, it is important to consider the time window the present technique is sensitive to. For example, Fig. 4c shows that doubling the diffusion constant from $0.3 \times 10^{-12}$ m²/s to $0.6 \times 10^{-12}$ m²/s leads to a significant change of the predicted signal envelope, which more quickly converges to the longer-lived tail produced by the adsorbed nuclei. In particular, molecules diffusing distances greater than 10 nm on a microsecond scale — corresponding to self diffusion coefficients of order $\sim 10^{-11}$ m²/s or greater — leave a negligible imprint on the correlation signal. Correspondingly, fluidic systems such as water ($D_{water} \sim 2.3 \times 10^{-9}$ m²/s) would be detectable only through the formation of a semi-mobile layer adjacent to the diamond surface. We emphasize that the sensitivity to the local system dynamics is not immediately indicative of accuracy in the derived parameters, which here must be understood as moderate given the crude assumptions of our model (spherical versus linear molecules, isotropic versus anisotropic diffusion, etc.; see Section E of the Supplementary Information). Along the same lines, we mention that it is difficult to ascertain whether the thin adsorbed layer observed in our experiments strictly originates from molecules in the fluid. In particular, recent studies have shown the presence of a protonated layer of comparable thickness possibly formed by water molecules or other hydrocarbons adsorbed upon ambient exposure[12, 13]. Controlled preparation of the diamond surface combined with NV-based NMR of spin species other than protons (e.g., [19]F or [31]P) can provide the means to more precisely separate contributions from sample molecules structured near the solid-liquid interface. Finally, deeper NVs (e.g., 15-20 nm from the surface) could be used to increase the relative contribution from more distant nuclear spins not adsorbed on the diamond surface, though at the expense of an overall lower detection sensitivity and spatial resolution as well as longer interrogation times[19].

On a final note, we hypothesize that the impact of the diamond crystal on the near-surface dynamics of the organic system is not restricted to fluidic samples but more in general, extends to most soft condensed matter systems. In particular, comparison between NV-detected and inductively detected NMR signals — omitted here for brevity — reveals longer coherence lifetimes for nuclear spins within the bulk of solid Samples A and B (200 μs and ~500 μs, respectively), which we interpret as indicative of a more restricted dynamics near the diamond crystal. Naturally, this conclusion relies on the connection between the nuclear spin coherence lifetime and the molecular mobility in these systems, a notion we confirm by examining control samples engineered to exhibit a variable degree of rigidity (Section G of the Supplementary Information).

In summary, the results herein introduce a new strategy to nanoscale nuclear spin sensing where the signal is recorded in the form of an 'FID' without the need for nuclear spin polarization or molecular labeling. The inherently small detection volume — of order 100 nm$^3$ in the present case — makes this form of sensing ideal to investigate dynamical processes on mesoscales, hard to access with other experimental techniques. For example, our approach could be exploited to more clearly expose the role of nanoscale surface roughness on the dynamics of flow, or to experimentally test differing boundary conditions invoked at the liquid-solid interface[23, 24] (e.g., no slip, multilayer locking, etc). Extensions articulating the present technique with known NMR protocols (e.g., homo- or heterospin decoupling sequences[26]) can be used to shed light on the chemical composition of adsorbed films in cases where the dynamics are insufficient to suppress inter-nuclear couplings. In particular, nuclear spin manipulation in the form of radio-frequency multipulse sequences can be applied during $\tilde{\tau}$ without deleterious effects on the NV response. Likewise, spin swap schemes — designed to exchange the spin states of the NV and its $^{14}$N (or $^{15}$N) host during the correlation interval [17] — provide a route to probe slow dynamical processes on time scales exceeding the NV longitudinal relaxation time[6]. Studies in this regime — more susceptible to the limited fluorescence contrast affecting the present correlation protocol — can benefit from the enhanced photon collection efficiency recently demonstrated for NVs within engineered diamond nanostructures[27, 28, 29].

More in general, the present technique opens interesting opportunities for the investigation of chemical or biochemical systems affected by compositional heterogeneities or local aggregation. Cell membranes in particular could serve as a fascinating research platform, since molecular diffusion is non-Brownian and mostly restricted to nanoscale fluidic pockets (presumably) separating more rigid structures[30]. By the same token, experiments as a function of temperature and/or the composition of the bulk fluid can help explore various thermally- or chemically-activated processes including, for example, protein folding or the dynamics of molecular motors tethered to the diamond surface.

## Acknowledgements

We thank Dr. D. Pagliero and Dr. Y. Li for assistance with some of the experiments. T.S.

acknowledges support from the IMPRS-AM. C.A.M acknowledges support from the National Science Foundation through grants NSF-1401632 and NSF-1309640. This work was supported by the German Science Foundation: SFB/TR21, FOR1694, EU SQUTEC, SIQS, and the Max Planck Society.

## Author contributions

N. R., S. P. and J. M. implanted the diamond sample for the preparation of shallow NV centers. C. A. M. and F. R. designed the experiment, and T. S. performed the experiments and analyzed the data. J. W. supervised the project. All authors contributed to the writing of the manuscript and gave approval for its final version.

**Probing molecular dynamics at the nanoscale via an individual paramagnetic center**

T. Staudacher[1,2], N. Raatz[3], S. Pezzagna[3], J. Meijer[3], F. Reinhard[1, 4], C. A. Meriles[5,*], and J. Wrachtrup[1]

[1]3rd physics Institute and Research Center SCoPE, University of Stuttgart, 70569 Stuttgart, Germany

[2]Max Planck Institute for Solid State Research, 70174 Stuttgart, Germany

[3]Jan Meijer

[4]TUM – Technical University of Munich, Walter Schottky Institut, Am Coulombwall 4, 85748 Garching, Germany

[5]Department of Physics, CUNY-City College of New York, 160 Convent Ave, New York, NY 10031, USA

*For correspondence: cmeriles@sci.ccny.cuny.edu

## A. Materials and Methods

### Diamond samples

We use Type IIa CVD grown (100)-oriented diamond samples from Element6. The NV centers used in this study were generated via 2.5keV $^{15}N^+$ ion implantation into the diamond substrates. The substrates were subsequently annealed at 240°C for 2h, followed by an 8h annealing at 850°C temperature, both in high vacuum. Afterwards the diamonds were boiled in a 1:1:1 mixture of $H_2SO_4$, $HClO_4$ and $HNO_3$, in order to remove any residual graphitic contaminations from the diamond surfaces.

### Sample preparation

We measured the correlation signals for three different organic samples, two organic polymers (A, B), and a liquid sample (C).

For the measurements of the solid compounds we spin coated a visibly thick layer of the respective material onto the diamond surface. Sample A is a "liquid coverslide" material dissolved in toluene (Merckoglas, Merck), which has a refractive index similar to glass and immersion oil. We dilute the Merckoglas base solution with toluene in a 1:1 ratio prior to spin coating. For the analysis we treat Merckoglas as a typical organic polymer, such as Poly(methyl methacrylate) (PMMA), and assume a comparable proton density.

Sample B is a film of Polydimethylsiloxane (PDMS). For this purpose a droplet of PDMS (Sylgard 184 Silicone Elastomer, Dow Corning in a 10:1 mixing ratio of the base to the curing agent) was spin-coated onto the diamond surface. Afterwards the PDMS is annealed by placing the diamond onto a hotplate for 2h at 80°C. We note that the signal to noise ratio under the PDMS coating is much smaller than for the other two samples, and requires longer measurement times. This is due to the larger mismatch of refractive indices, and the accompanied loss of fluorescence signal.

The measurements of the liquid sample C were performed by covering the diamond surface with immersion oil (Fluka Analytical, 10976).

Before/after the measurements, the samples were removed by rinsing the diamond in a solvent while sonicating. The diamond is subsequently boiled in the above-mentioned 1:1:1 acid mixture for multiple hours to remove any organic residues of the sample or solvent from the surface.

To ensure the observed NMR signal (mainly) originates from the sample rather than from adsorbates due to exposure to the ambient environment, the diamond is removed from the acid and immediately covered with the respective sample, so as to minimize the exposure time to the ambient.

## B. Signal reproducibility

All experiments were carried out with the same diamond sample, and thus with the same set of shallow NVs. However, because the diamond crystal must be physically removed from the microscope for surface coating, it is difficult to use the same individual NV for testing nuclear spins from different sample films. To circumvent this complication, we collected data from multiple (50 to 100) individual NVs exposed to the same sample film (see below). Among these only a few can be dynamically decoupled for a time sufficiently long to see the nuclear spin signature via an XY8-N sequence (a likely result of the NV depth dispersion and heterogeneity of the local concentration of paramagnetic impurities). We find that not every NV center exhibiting a nuclear spin induced dip under the XY8-N sequence also shows detectable nuclear spin correlation. The reason for this is still not fully understood but it may relate to the fortuitous overlap between the proton spin signature and the fourth harmonic of the $^{13}$C-induced dip (see below, Section H): The latter can be easily mistaken by the former in an XY8-N sequence but not in the correlation protocol (where the nuclear spin Larmor frequency is directly probed). Also worth noting is the small relative amplitude of the correlation signal ($\approx 10\%$ of the maximum possible fluorescence contrast between the NV spin states), which leads to a correspondingly low SNR and makes detection difficult. Within these experimental limitations, Figs. 2 and 3 in the main text show *representative* data sets from the NVs that did display a sizeable correlation signal. Among the latter, excellent reproducibility was observed between data sets in all the samples we explored, as shown immediately below.

## Sample A – Merckoglas

We examined a set of 52 NV centers, out of which 10 could be dynamically decoupled to coherence times of $\approx 100$ μs, allowing us to detect the proton signal via an XY8-10 sequence. Only 4 of these NVs show a detectable signal after an XY8-3 correlation sequence.

Fig. S1 shows three of those NV centers, the one presented in the main text (NV20) and two additional ones. We find a reproducible correlation signal, described by an exponentially damped sinusoid centered around the proton Larmor frequency. The decay envelope (here maintained unchanged for all NVs in the figure) is qualitatively captured by the presented model, with a decay time of 20 μs.

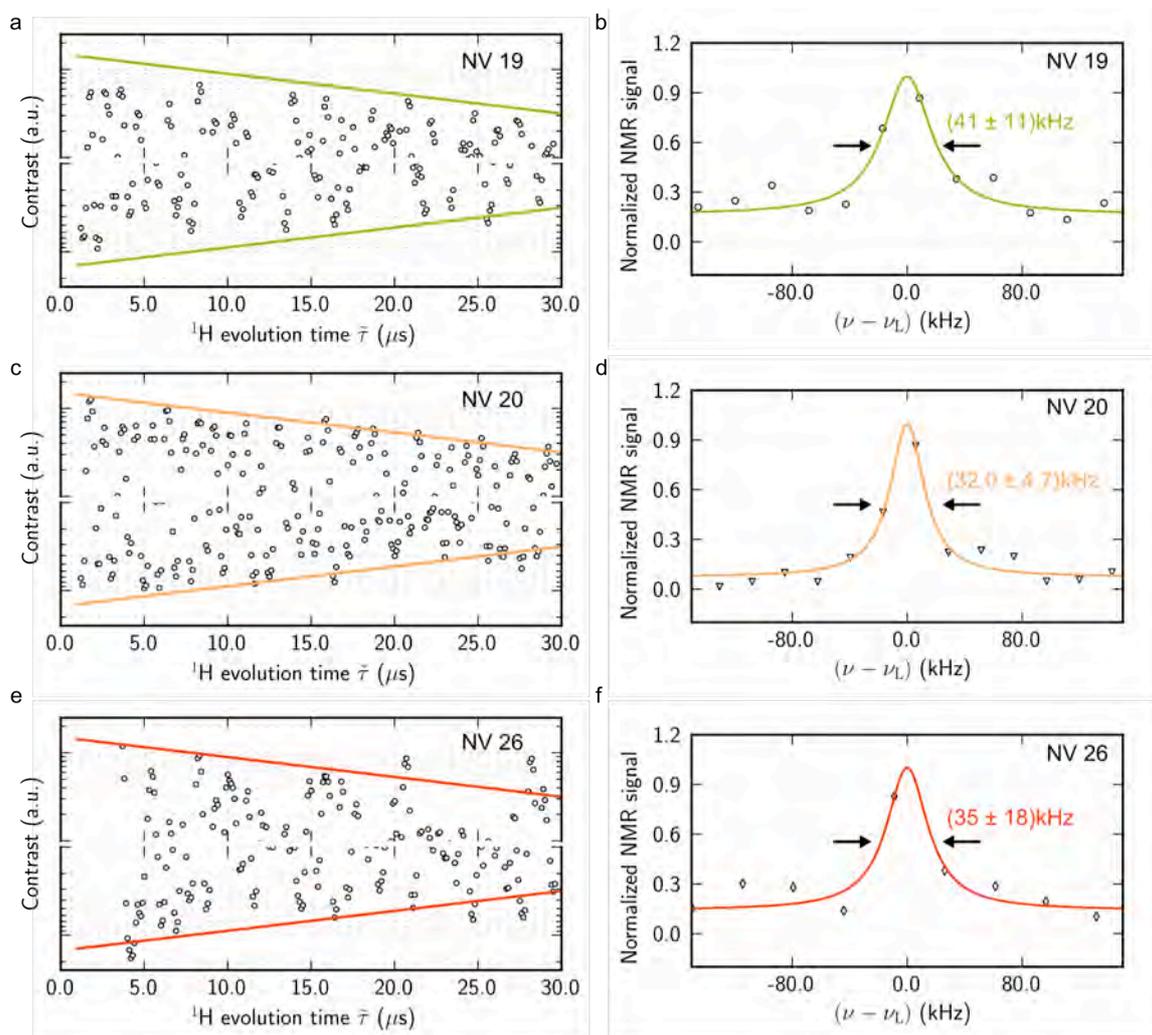

**Figure S1. Signal reproducibility for Sample A.** (a) Proton correlation signal as detected from an NV center referred to as NV 19 (open circles) for Sample A (Merckoglas). The solid trace is the envelope of our model adapted to a fully static proton bath. (b) Fourier transform (magnitude mode) of the signal in (a) centered at the proton Larmor frequency $\nu_L$ along with a Lorentzian fit (solid green trace). (c,d) Same as in (a,b) for NV 20. (e,f) Same as in (c,d) but for NV 26.

## Sample B – PDMS

For the PDMS coating we examined a set of 85 NV centers, out of which 13 showed proton signal via an XY8-10 sequence. Only 2 of these 13 NVs reveal a noticeable signal after an XY8-3 correlation sequence.

The two correlation curves shown in Fig. S2 show the curve presented in Fig. 3 of the main text (NV1) and a second curve for a second NV (NV78). As explained above, experiments with this sample suffered from poor collection efficiency and required particularly long integration times, the reason why the second data set was measured only for evolution times < 20 μs. Both measured signals are overlayed with the same calculated decay envelope, assuming a slow molecular dynamics of the PDMS molecules as described in the main text (see also Section D below).

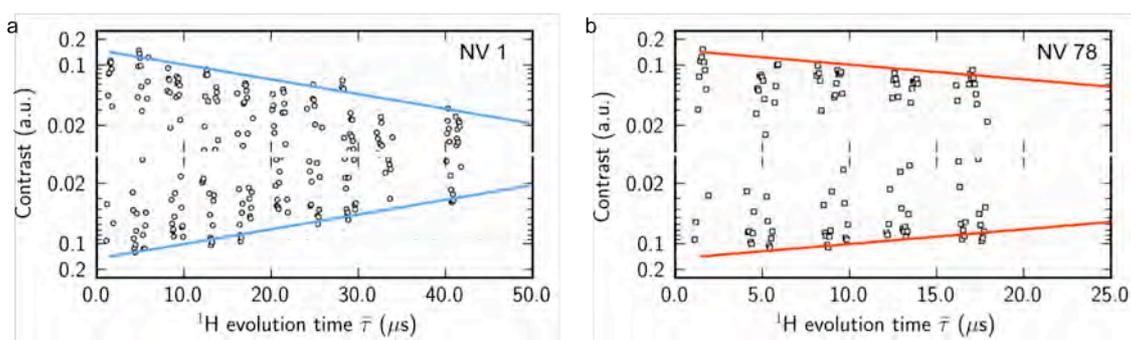

**Figure S2. XY8-3 correlation signal under a coating of sample B for two different NV centers.** The time domain data is compared with a decay envelope calculated assuming partial motional narrowing due to slow molecular dynamics as described above.

## Sample C – Immersion oil

For the immersion oil a set of 83 NV centers was measured. Among them, nine NVs exhibited proton signal after an XY8-10 sequence. Five of those NVs showed an XY8-3 correlation signal. We measured two of those NVs over a longer evolution time window (> 80µs, NV6 is the one shown in the main text and related SI sections). As shown in Fig. S3, we measure virtually identical responses characterized by a fast initial decay and a long-lived tail as highlighted in the main text. Remarkably, the two data sets are so consistent that the same set of parameters (adsorbed layer thickness, translational diffusion constant, and rotational diffusion constant) can be used to describe both observations.

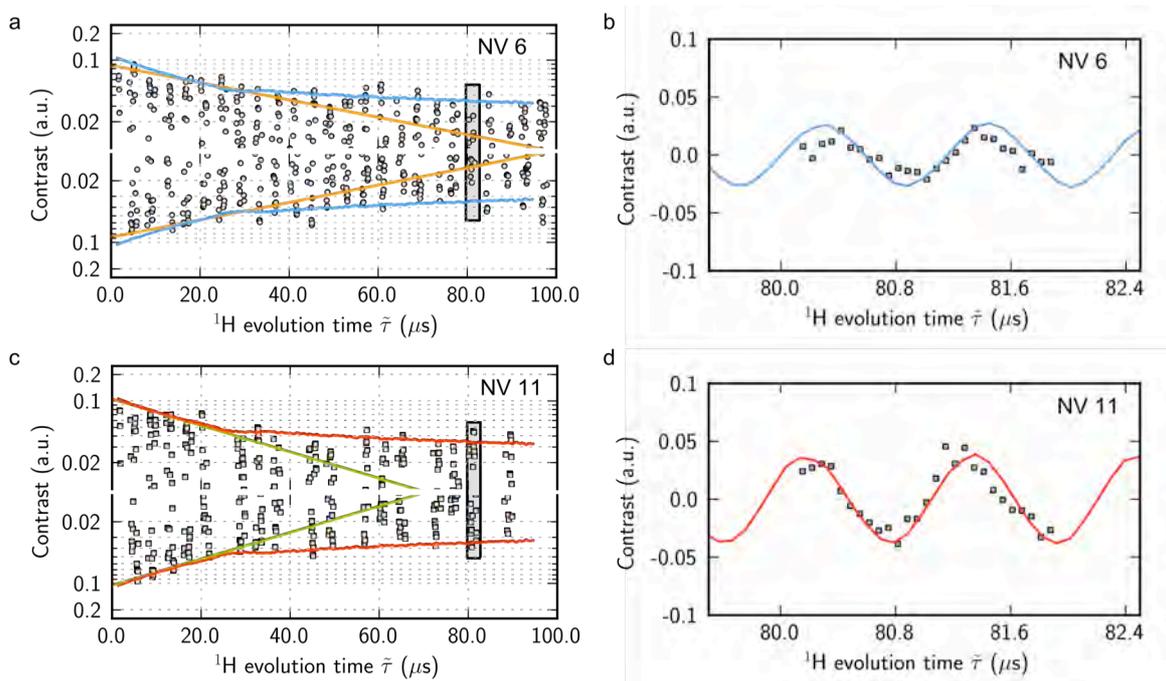

**Figure S3. Signal reproducibility in the presence of a liquid.** (a) Proton correlation signal from a shallow NV (referred to as NV 6, open circles) in the presence of immersion oil. The yellow and light blue traces are the envelopes of an exponentially decaying sinusoidal and the presented model, respectively. The exponential fit considers the initial decay $\tilde{\tau} \leq 25$ µs whereas the model considers contributions from molecules in the bulk liquid and semi-mobile molecules from an adsorbed layer as described in the main text. (b) Zoomed image of the data points within the shaded area in (a); the light blue trace is the segment of a fit to a sinusoidal function at the proton Larmor frequency scaled by the envelope in (a). (c,d) Same as in (a,b) but for a different NV (identified as NV 11 in the image). In both cases we use the same parameters, namely, the adsorbed film thickness as well as the translational and rotational diffusion coefficients.

## C.  Modeling the correlation signal envelope

The envelope of the correlation signal is described by (see Eq. (2) of the main text)

$$\sum_{i,j} p_{i,j}(\tilde{\tau}) \, e^{-\tilde{\tau}/T_2^{i,j}}, \qquad (S1)$$

where $p_{i,j}(\tilde{\tau})$ is the conditional probability that the $i$-th nuclear spin of the $j$-th molecule remains in the detection volume over the evolution time $\tilde{\tau}$, and $T_2^{i,j}$ is the transverse relaxation time of said nuclear spin. Here we neglect any additional contribution to the envelope arising from longitudinal relaxation of the NV electron spin (see below Section F). This is a reasonable assumption as typical longitudinal relaxation times are in the order of hundreds of μs, while we focus here on effects on timescales of a few tens of μs, i.e. the time range of our measurements. Before describing how we model the probability and relaxation time in detail, we introduce below the underlying assumptions and simplifications of the presented theory.

First, we assume that the value of the sum in Eq. (S1) is time independent, i.e. the total number of protons/molecules inside the detection volume stays approximately constant over time. Therefore the amplitude and decay components of the correlation signal depend solely on the spatial distribution of the molecular mobility with respect to the surface distance. Further, we neglect any hydrodynamic interactions among molecules other than those implicit when assigning a candidate translation or rotation diffusion coefficient.

The molecules are treated as rigid spheres of radius $a$, and their molecular mobility varies on the length scale of the detection volume. This heterogeneity is described by position dependent diffusion constants $D_{\mathrm{T}}(\vec{r})$ and $D_{\mathrm{R}}(\vec{r})$ for the translational and rotational diffusion respectively, which takes into account the possible presence of adsorbate molecules on the diamond surface. We estimate that the transition between adsorbate and free molecules is sharp and that the latter obey classical diffusion equations, as well as the Stokes-Einstein-(Debye) equations for the diffusion constants

$$D_{\mathrm{T}} = \frac{k_{\mathrm{B}} T}{6 \pi \eta a} \; ; \qquad (S2)$$

$$D_R = \frac{k_B T}{8\pi\eta a^3} \ ; \qquad\qquad (S3)$$

where $k_B$ is the Boltzmann constant, $T$ is the medium temperature, and $\eta$ is the viscosity of the sample medium.

We further assume that the diffusion heterogeneity is restricted to the z-axis perpendicular to the diamond surface. The rotational diffusion constant can have a finite value at the surface, while the translational diffusion constant goes to zero. These values stay approximately constant until a distance $z_{ads}$ to the surface, which we define as the thickness of the adsorbate layer. Beyond this layer, the diffusion constants are related to each other via $D_R = \frac{3}{4a^2} D_T$. We assign to both diffusion constants a similar transition profile, which we model as (Fig. S4a)

$$D(z) = \begin{cases} D_{surface}, \ z \le z_{ads} \\ (D_{bulk} - D_{surface})\left(1 - \exp\left(-\frac{z - z_{ads}}{\zeta}\right)\right) + D_{surface}, \ z > z_{ads} \end{cases}, \qquad (S4)$$

where $D_{bulk}$, and $D_{surface}$ are the respective values of the diffusion constants in the bulk of the volume and at the surface, and $\zeta$ is a transition parameter, which we keep fixed at an empirical value of $\zeta = 0.25$ nm. We note that this value is in the same order of magnitude as that found for AFM experiments measuring the transition of surface to bulk fluid dynamics[1]. We find that the exact value of $\zeta$ has a negligible impact on the results of the simulation, as long as the transition takes place on a scale small compared to the radial size of the detection volume.

The coherences of the nuclear spin bath characterized by the transverse relaxation time $T_2^{i,j}$ can be calculated using established means known from classical NMR[2, 3, 4]. In the following we will briefly outline the calculation steps of the relaxation time, before describing the model we use to estimate the conditional probability $p_{i,j}(\tilde{\tau})$.

## Nuclear spin relaxation time $T_2$

The nuclear spin dynamics mainly depends on position dependent correlation times of motion $\tau_i$.

The only interaction considered among the nuclei is homonuclear dipolar coupling, so that one can distinguish different time regimes by the ratio of the rate of nuclear spin reorientation ($\propto \tau_i^{-1}$) versus the dipolar coupling constant $\omega_D$.

Among the interactions responsible for the correlation decay one can distinguish between intramolecular interactions inside a molecule, and intermolecular interactions among spins of different molecules. In the interior of a molecule, the variation of the dipolar coupling between spins arises almost exclusively from the rotation of the molecule (neglecting distance variations due to vibrations), while for the interactions between spins in different molecules their relative translation must be considered.

### 1H-1H interaction

The homonuclear dipolar coupling between two proton spins is described by the Hamiltonian

$$H_{\text{dip}} = \frac{\mu_0 \hbar^2}{4\pi} \frac{\gamma_{1H}^2}{r_{HH}^3} \left( I_1 I_2 - 3 \frac{(\vec{r}_{HH} I_1)(\vec{r}_{HH} I_2)}{\vec{r}_{HH}^2} \right), \qquad \text{(S5)}$$

where $\mu_0$ is the magnetic constant, $\hbar$ is the reduced Planck constant, $\gamma_{1H}$ is the gyromagnetic ratio of a proton spin, $I_i$ are the spin vector operators of the i-th proton, and $\vec{r}_{HH}$ is the distance vector between two proton spins. Squaring Eq. (S5), averaging over the azimuthal and polar angles, and taking the trace over the spin matrices yields an expression for the rms-dipolar coupling strength[5, 6]

$$\sqrt{\omega_D^2} = \frac{63}{8} \sqrt{\frac{3}{5}} \frac{\mu_0 \hbar}{4\pi} \frac{\gamma_{1H}^2}{r_{HH}^3}, \qquad \text{(S6)}$$

where we use a representation of $\vec{r}_{HH}$ in spherical coordinates. In Eq. (S6) $r_{HH}$ is the average distance to the nearest nuclear neighbor, which is dependent on the proton density $\rho$ and is distributed according to

$$P(r_{HH}) = 4\pi\rho \, r_{HH}^2 \exp\left(-\frac{4\pi}{3} r_{HH}^3 \rho\right) \, . \qquad \text{(S7)}$$

This results in an average proton-proton distance of

$$\langle r_{\mathrm{HH}} \rangle \approx \frac{1}{2}\, \rho^{-\frac{1}{3}} . \qquad (S8)$$

Combining Eq. (S8) and Eq. (S6) yields an expression for the average interaction strength

$$\sqrt{\omega_{\mathrm{D}}^2} = 63\, \sqrt{\frac{3}{5}}\, \frac{\mu_0 \hbar}{4\pi}\, \gamma_{1\mathrm{H}}^2 \rho . \qquad (S9)$$

The relaxation rate of the axial magnetization component for a 2-spin cluster due to their mutual dipolar interaction is calculated via[5, 6, 7]

$$f_{\mathrm{dip}} = \frac{3}{2}\, \frac{\mu_0 \hbar}{4\pi}\, \gamma_{1H}^2\, \frac{4\pi^2}{3}\, \rho \quad . \qquad (S10)$$

Since we assume a constant proton density throughout the detection volume, the average interaction strength and dipolar relaxation rate are therefore position independent. Using a typical value for the proton density in organic compounds of $\rho = 50\ \mathrm{nm}^{-3}$ for all samples, yields a fluctuation rate of $f_{\mathrm{dip}} \approx 20\ \mathrm{kHz}$, and average interaction strength of $\sqrt{\omega_{\mathrm{D}}^2} \approx 45\ \mathrm{kHz}$.

### *Translational diffusion of 1H spins*

The translational diffusion of molecules generates a small fluctuating magnetic field. The fluctuation rate due to translational diffusion is given by[5]

$$f_{\mathrm{trans}}(\vec{r}) = 2D_{\mathrm{T}}(\vec{r})\, \left(\frac{1}{2\,a}\right)^2 . \qquad (S11)$$

### *Rotational diffusion of 1H spins*

For a spherical molecule of radius $a$ rotating in a liquid of viscosity $\eta$ the rotational fluctuation rate is described by Stokes' law[2]

$$f_{\mathrm{rot}}(\vec{r}) = \frac{3\, k_{\mathrm{B}} T}{4\pi a^3\, \eta(\vec{r})} . \qquad (S12)$$

Using the Stokes-Einstein relation (Eq. ( S3)) we find

$$f_{\mathrm{rot}}(\vec{r}) = 6D_{\mathrm{R}}(\vec{r}) . \qquad (S13)$$

### Combined dynamics and relaxation rate

In order to describe the nuclear spin relaxation dynamics of the sample molecules we use an autocorrelation function of the fluctuating magnetic field B(t) experienced by the nuclear spins

$$\langle B(t)B(t')\rangle = \langle B^2 \rangle \exp(-\frac{|t-t'|}{\tau_c}), \qquad (S14)$$

where $\langle B^2 \rangle$ is the rms field strength experienced by the protons spins, and $\tau_c$ is the characteristic correlation time of the magnetic fluctuations, which includes all above mentioned mechanisms, i.e. $\tau_c(\vec{r}) = (f_{dip} + f_{trans}(\vec{r}) + f_{rot}(\vec{r}))^{-1}$. The spectral density $\tilde{J}(\omega, \tau_c)$ is twice the Fourier transform

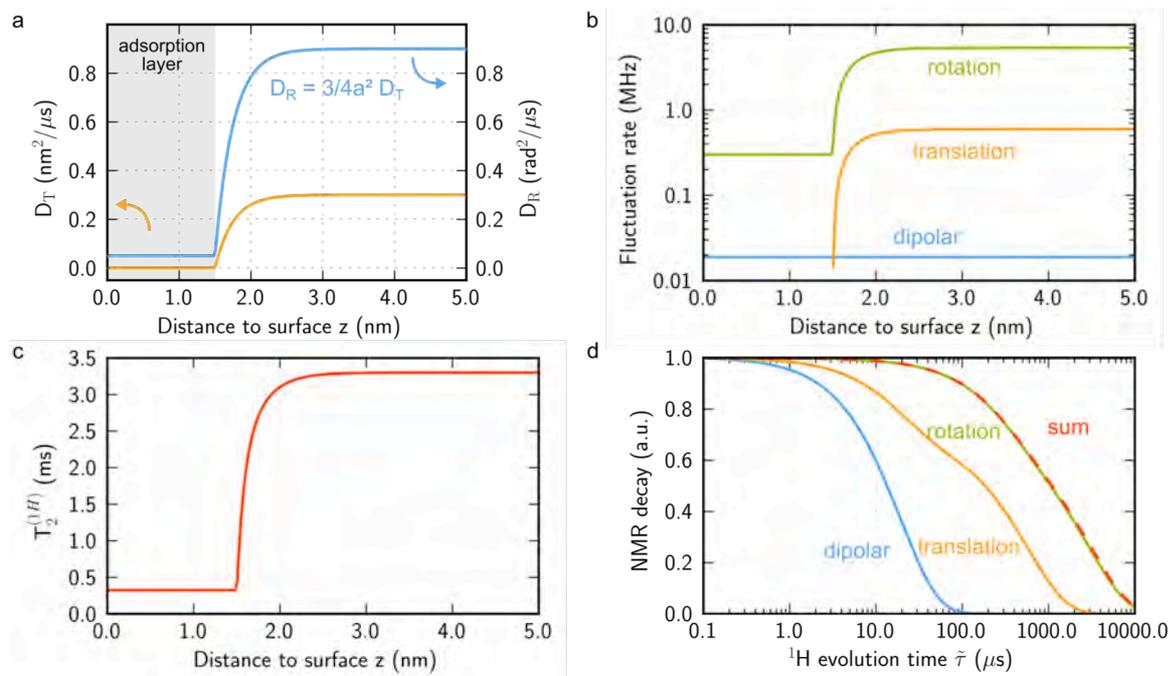

**Figure S4. Calculation of the nuclear spin relaxation process.** (a) Translational and rotational diffusion constants with increasing distance to the diamond surface. The rotational constant has a finite value at the surface, and both constants are linked via the Stokes-Debye-Einstein relations in the outside layer. (b) The varying diffusion constants result in different fluctuation rates throughout the detection volume, the dipolar interaction is constant, as we assume a constant proton density throughout the volume. (c) The relaxation time is dominated by the strong dipolar interactions in the static adsorption layer but quickly prolongs with faster molecular dynamics. (d) Integrated NMR decay over the detection volume, for different molecular interactions. The NMR signal quickly decays in the absence of molecular dynamics (blue curve), gets longer in the presence of translational diffusion (yellow curve), and is maximum if also rotational diffusion is allowed (green, red curve). Same parameters as in the main manuscript: molecule radius of a = 0.5 nm, proton density $\rho$ = 50 nm$^{-3}$, adsorption layer thickness $z_{ads}$ = 1.5 nm, surface translational diffusion constant of $D_{T,surface}$ = 0.0 nm$^2$/μs, bulk translational diffusion constant of $D_{T,bulk}$ = 0.3 nm$^2$/μs, a rotational diffusion constant on the surface of $D_{R,surface}$ = 0.05 rad$^2$/μs, and a rotational diffusion constant in the bulk of $D_{R,bulk}$ = 3/4a$^2$ $D_{T,bulk}$.

of Eq. (S14) [4]

$$\tilde{J}(\omega, \tau_c) = 2\langle B^2 \rangle \frac{\tau_c}{1 + \omega^2 \tau_c^2}. \qquad (S15)$$

The normalized spectral density $J(\omega, \tau_c)$ is

$$J(\omega, \tau_c) = \frac{\tau_c}{1 + \omega^2 \tau_c^2}. \qquad (S16)$$

The transverse relaxation is caused by the dephasing of the proton spins due to their mutual dipolar interactions. In the case of intramolecular (homonuclear) dipole-dipole relaxation, the transverse relaxation rate $T_2^{-1}$ between two spin ½ nuclei is given by[3, 4]

$$T_2^{-1}(\vec{r}) = \frac{3}{20} \sqrt{\omega_D^2}^2 \left[ 3J(0, \tau_c(\vec{r})) + 5J(\omega_L, \tau_c(\vec{r})) + 2J(2\omega_L, \tau_c(\vec{r})) \right], \qquad (S17)$$

where $\omega_L$ is the nuclear Larmor precession frequency, and $\sqrt{\omega_D^2}$ is the average dipole-dipole coupling constant, given by Eq. (S6).

## Probability $p_{i,j}(\tilde{\tau})$

We model the probability $p_{i,j}(\tilde{\tau})$ using a semiquantitative geometric model. Simply put, we draw a 'diffusion sphere' around a starting position $\vec{r}$ with a radius

$$r_{diff}(\vec{r}, \tilde{\tau}) = \sqrt{6 D_T(\vec{r}) \tilde{\tau}}, \qquad (S18)$$

which is the rms-diffusion distance in three dimensions for Brownian motion. We only consider translational diffusion here, and the translational diffusion constant $D_T(\vec{r})$ only depends on the starting position $\vec{r}$ (Fig. S4a, Fig. S5a). The conditional probability to find the molecule inside the detection volume ($V_{DV}$) after the time $\tilde{\tau}$ is then approximately given by the ratio of the overlap between the diffusion volume ($V_{diff}$) and the detection volume, normalized to the detection volume (Fig. S5b):

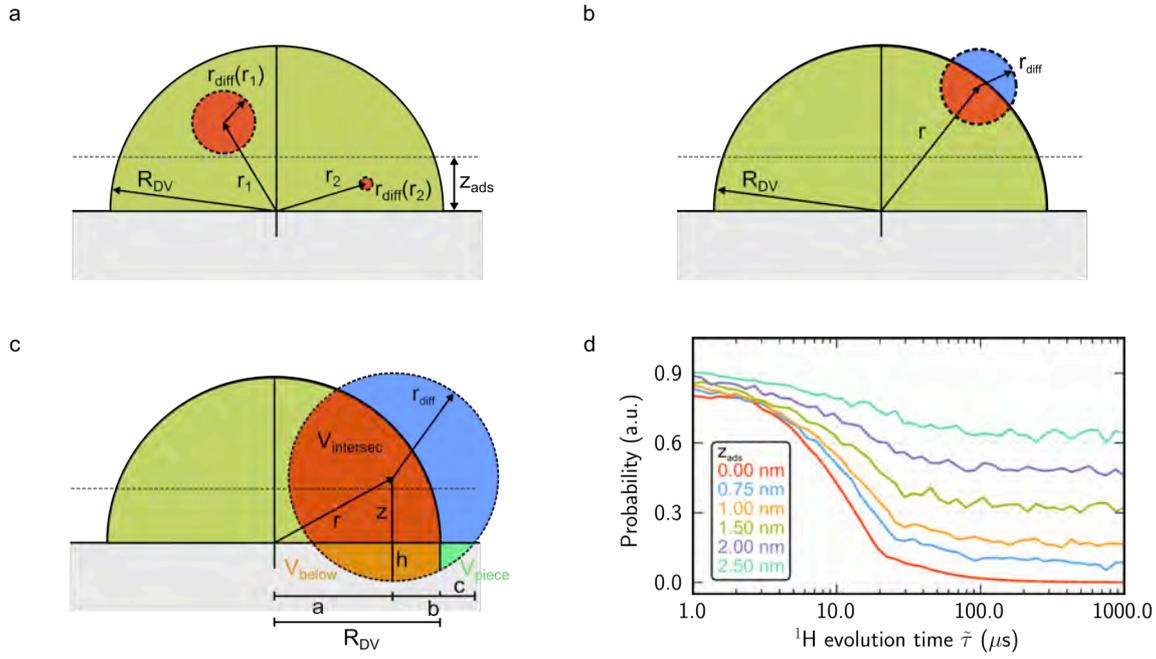

**Figure S5. Geometric model for signal loss due to translational diffusion** (a-c) The detection volume (green) can be approximately described by a heterogeneous hemisphere comprising a static near-surface layer and a mobile outer segment. The probability of a molecule staying within the detection volume during $\tilde{\tau}$ depends on the assumed translational diffusion constant in the outer section of the hemisphere. This probability (growing from (a) to (c)) is described by the overlap (red) of the diffusion sphere with the detection volume. (d) For a translational diffusion constant of $D_{T,bulk}$ = 0.3 nm²/µs, we find that the probability of molecules staying inside the detection volume quickly diminishes (red curve), unless a sizeable adsorption layer is present.

$$p_{i,j}(\tilde{\tau}) \ = \ \frac{V_{\mathrm{diff}} \cap V_{\mathrm{DV}}}{V_{\mathrm{DV}}}. \tag{S19}$$

The detection volume is simplified as a hemisphere with radius $R_{\mathrm{DV}} = 4$ nm, which roughly translates to an expected volume of (5 nm)³.

The overlap $V_{\mathrm{diff}} \cap V_{\mathrm{DV}}$ can be calculated via the intersection of two spheres

$$V_{\mathrm{intersec}} \ = \ \frac{\pi \, (R_{\mathrm{DV}} + r_{\mathrm{diff}} - r)^2 (r^2 + 2r(R_{\mathrm{DV}} + r_{\mathrm{diff}}) - 3(R_{\mathrm{DV}} - r_{\mathrm{diff}})^2)}{12\, r}, \tag{S20}$$

where $r$ is the absolute value of the starting position vector $\vec{r}$, and $r_{\mathrm{diff}}$ is obtained from via Eq. (S18). The intersection volume will have a negative value if either the diffusion sphere is completely inside the detection volume ($p_{i,j}(\tilde{\tau}) \ = 1$), or if the diffusion sphere becomes larger than the detection volume. In the latter case the probability $p_{i,j}(\tilde{\tau})$ is determined by the ratio of the diffusion sphere volume to the detection volume.

Since the detection volume is only a hemisphere, we have to subtract any intersection or diffusion volume contributions extending below the surface before calculating the value for the probability. The volume beneath the surface can be calculated as the volume of a sphere cap ($V_{\text{cap}}$). Therefore the volume below the surface can be calculated as

$$V_{\text{below}} = V_{\text{cap}} - V_{\text{piece}}, \qquad (S21)$$

with $V_{\text{cap}} = \frac{\pi}{3} h^2 \left(3 r_{\text{diff}} - h\right)$, where $h = r_{\text{diff}} - z$ is the height of the cap, and $z$ is the z-component of the position vector $\vec{r}$. The volume of the piece which would be subtracted is approximately given by

$$V_{\text{piece}} \approx \frac{c}{2b} \, V_{\text{cap}}, \qquad (S22)$$

where $b = \sqrt{r_{\text{diff}}^2 - z^2}$ is the radius of the circular base of the cap volume, and $c = a + b - R_{\text{DV}} = \sqrt{r^2 - z^2} + \sqrt{r_{\text{diff}}^2 - z^2} - R_{\text{DV}}$.

## D. Monte Carlo (MC) simulation of the molecular dynamics in the detection volume and comparison to the experimental data

For the simulation we pick random starting positions $\vec{r}$ inside the detection volume, then calculate the probability to remain within this volume for varying $\tilde{\tau}$ as described above, i.e. varying the radii of the diffusion sphere. For each $\tilde{\tau}$ we calculate an average correlation time $\langle T_2^{(i,j)} \rangle$ between the highest and the lowest z-value in the diffusion sphere. The probability is then multiplied by the NMR decay $\exp(-\tilde{\tau} / \langle T_2^{(i,j)} \rangle)$. We do an ensemble average over multiple starting positions to get the envelope of the correlation signal. We vary the values of the adsorption layer thickness $z_{\text{ads}}$, the rotational diffusion constant at the surface $D_{\text{R,surface}}$, and the translational diffusion constant in the bulk $D_{\text{T,bulk}}$, and overlay the results with the experimental data to obtain quantitative estimates of the molecular mobility in the samples.

We find good qualitative agreement between the simulation and the experimental data. For Sample C (immersion oil, Fig. S6) the model manages to reproduce both significant features, the initial fast decay (attributed to the molecules diffusing outside the detection volume), and the long lasting signal tail (described by the NMR decay of the molecules inside the adsorption layer). The 'amplitude' of the long lasting decay is set by the thickness of the adsorption layer (Fig. S5d), from which a value of 1.5 nm is estimated. This result is in good agreement with prior experimental observations[8, 9], and would correspond to approximately 1-2 layers of molecular adsorbates on the surface. The translational diffusion constant mainly influences the fast initial drop of the signal, and we find best agreement for a value of $D_{T,bulk} \sim 0.3$ nm$^2$/µs. We note that this is in the order of magnitude that one would expect from Eq. (S2), which yields a value of $D_{T,bulk} \sim 1.0$ nm$^2$/µs, corresponding to a molecule of radius $a$ = 0.5 nm, a medium of viscosity $\eta$ = 437 mPa.s (as listed in the compound datasheet), and room temperature conditions. The deviations could be readily explained by varying the molecule radius a, as the immersion oil contains various different molecules of different chain length etc., which are treated identically in the simulations.

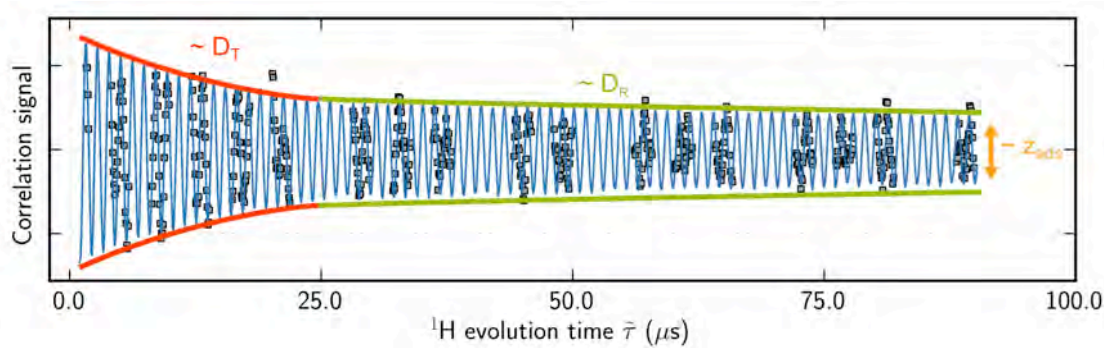

Our model also manages to describe the decay times of the purely exponential signal envelopes for the solid polymers quite well. Here we assume no translational diffusion throughout the detection volume, i.e. the whole volume is treated as an 'adsorption layer'. We then vary the value of the rotational diffusion constant $D_{R,surface}$ until we obtain a reasonable match with the experimental data.

In the case of Merckglas (sample A) we can only reproduce the signal shape by removing the rotational diffusion as well, i.e. molecules remain static over the course of the measurement. In this case the probability $p_{i,j}(\tilde{\tau})$ becomes unity, and the signal decay is mainly governed by NMR relaxation due to inter- and intramolecular dipolar interactions, yielding a fast single exponential decay (Fig. S7a) as expected for a solid.

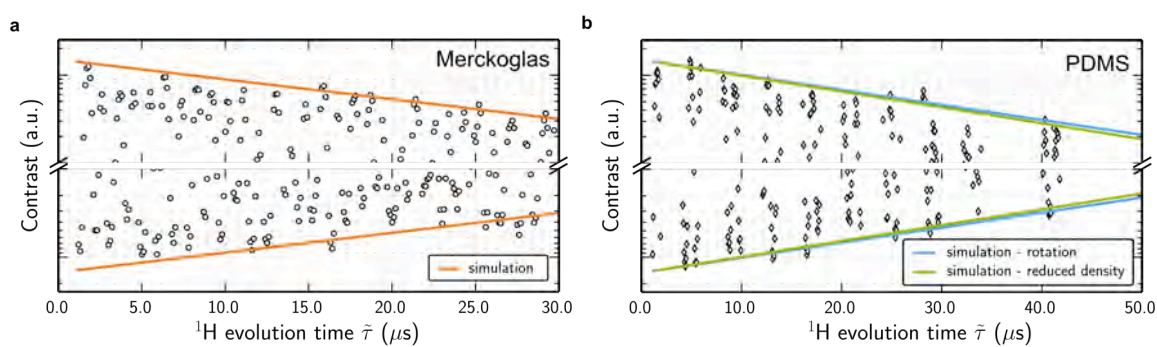

**Figure S7. Simulated correlation envelopes for the solid-state samples** (a) The Merckglas data can be readily described by a completely static bath of molecules, i.e. by a purely dipolar NMR relaxation, yielding a single exponential decay envelope (yellow curve), in good agreement with Eq. 1 in the main manuscript. (b) The PDMS data can be readily reproduced by either assuming slow molecular rotation with no center-of-mass translation (blue curve, $D_{R,surface}$ = 0.001 rad$^2$/μs), or by assuming a completely static layer with a (~20%) reduced proton density of 40 nm$^{-3}$ (green curve).

For PDMS the correlation signal also decays exponentially but the decay time (of order ~30 μs) is found to be longer than anticipated for an ensemble of static, dipolarly-coupled protons (at least assuming a standard proton density of $\rho$ = 50 nm$^{-3}$). We can reproduce the observed behavior by assuming a rotational diffusion constant of $D_{R,surface}$ = 0.001 rad$^2$/μs throughout the detection volume. The latter is supported by prior NMR experiments showing that PDMS can exhibit significant molecular dynamics at room temperature, thus leading to partial motional narrowing[10, 11]. The self-diffusion depends on the PDMS chain length/molecular weight, i.e., on the mixing ratio of the components. Even if partly, this enhanced mobility can be the reason of the line narrowing captured in the non-negligible value of the rotational diffusion constant in the simulation. For completeness, we note that the observed data set can also be reproduced by a static ensemble of protons with a ~20% lower density than normal (40 nm$^{-3}$ as opposed to 50 nm$^{-3}$, lighter green trace in Fig. S7b). However, in light of the experimental evidence mentioned above, this alternative seems less realistic (see also Section G).

## E. Model precision and accuracy

As described above the key parameters in our model are the bulk translational diffusion coefficient $D_{\text{T,bulk}}$ (which governs the initial signal decay for a liquid sample), the rotational diffusion coefficient in the semi-mobile adsorption layer $D_{\text{R,surface}}$ (describing the decay time of the long lasting signal tail), and the thickness of the adsorption layer $z_{ads}$ (which sets the amplitude of this long signal component). Fig. S6 schematically visualizes the role of each parameter on the resulting decay envelope.

As shown in Fig. S8 for Sample C, we find that the modeled response of NV 6 is quite sensitive to the values assigned to these parameters. For example, Fig. S8a shows the initial decay of the correlation signal due to the translational diffusion of the sample molecules for varying translational diffusion coefficients $D_{\text{T,bulk}}$. Here the values for the rotational diffusion inside the semi-mobile layer ($D_{\text{R,surface}} = 0.05 \text{ rad}^2/\mu s$) and the adsorption layer thickness ($z_{\text{ads}} = 1.5 \text{ nm}$) are kept constant. One can qualitatively estimate a confidence interval (light blue shaded area) of $D_{\text{T,bulk}} = [0.2 \frac{\text{nm}^2}{\mu s}, 0.4 \frac{\text{nm}^2}{\mu s}]$ for which the model parameters reproduce the envelope shape to a comparable extent. Beyond this interval the calculated response notably departs from the measurement, allowing us to distinguish changes of the diffusion coefficient within ~ 30%.

Fig. S8b shows the long decay component of the correlation signal for varying rotational diffusion coefficients $D_{R,surface}$. Here we keep the values for the translational diffusion in the bulk layer ($D_{T,bulk} = 0.3 \ nm^2/\mu s$) and the adsorption layer thickness ($z_{ads} = 1.5 \ nm$) constant. Over the measured proton evolution time we find that the correlation amplitude stays approximately constant, with no distinct signal decay. Therefore we can only estimate a lower bound for the rotational diffusion coefficient for which we can reproduce this behavior. We find that this signal shape is reproduced by rotational diffusion coefficients larger than $D_{R,surface} = 0.05 \ \frac{rad^2}{\mu s}$, and that a notable deviation from the data can be observed for coefficients below $D_{R,surface} = 0.02 \ \frac{rad^2}{\mu s}$. Therefore within the measurement region the model can distinguish changes below $0.05 \ \frac{rad^2}{\mu s}$ with a precision of $\sim 60\%$.

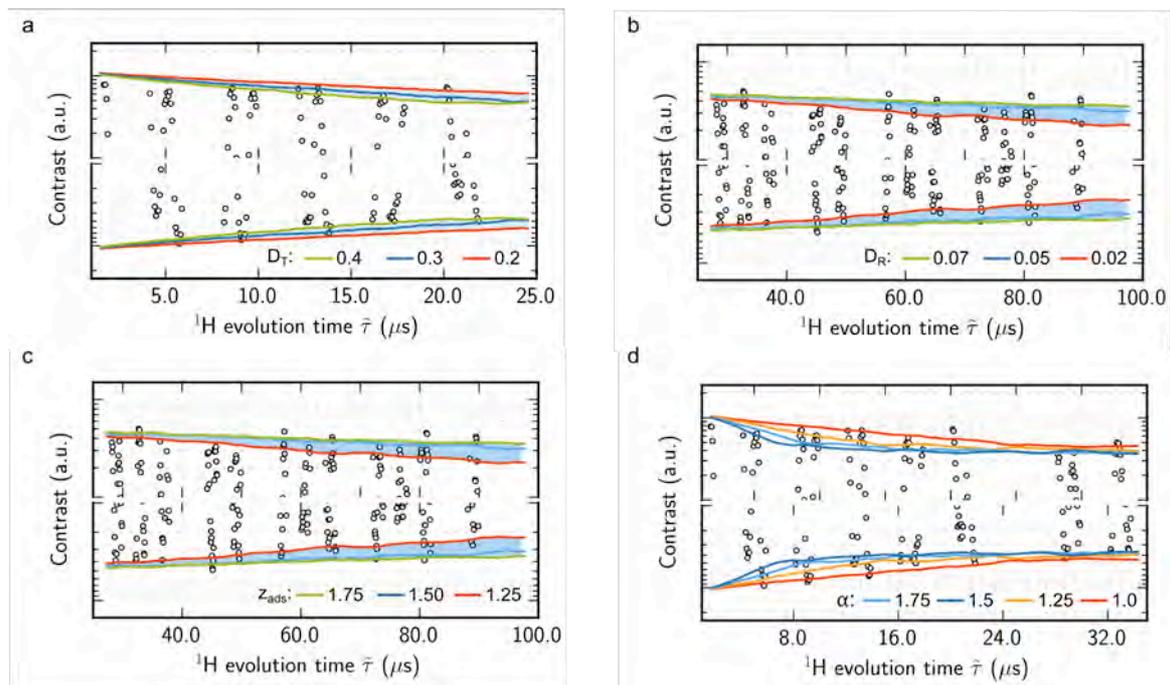

**Figure S8. Variation of the model parameters and influence in the obtained decay envelope.** (a) Variation of the translational diffusion coefficient $D_{T,bulk}$, the respective value is noted in units of [$nm^2/\mu s$]. (b) Variation of the rotational diffusion coefficient in the semi-mobile adsorption layer $D_{R,surface}$, the respective value is noted in units of [$rad^2/\mu s$].. (c) Variation of the adsorption layer thickness $z_{ads}$, the respective value is noted in units of [nm]. (d) Influence of anomalous diffusion on the Initial signal decay. In (a) through (d) the data points correspond to NV 6 and the diamond coating is sample C (immersion oil). In all cases only the noted parameter is changed while the others have a fixed value coincident with that noted in the main text.

Fig. S8c shows the variation in the amplitude of the long decay component of the correlation signal for varying adsorption layer thicknesses. In this case the translational diffusion in the bulk layer ($D_{\mathrm{T,bulk}} = 0.3\ \mathrm{nm}^2/\mu s$) and the rotational diffusion in the adsorption layer ($D_{\mathrm{R,surface}} = 0.05\frac{\mathrm{rad}^2}{\mu s}$) are kept constant. We can estimate a confidence interval of $z_{\mathrm{ads}} = [1.25\ \mathrm{nm},\ 1.75\ \mathrm{nm}]$, which results in comparable decay responses corresponding to a precision of $\sim 17\%$.

The relatively good precision of our model in discriminating between different numerical values of the fit parameters must be distinguished from the model absolute accuracy which, in light of the various simplifications (Section C), should be considered only moderate. For example, one of the underlying assumptions is that the sample molecules are spherical and that they diffuse according to a free Brownian motion. The effect of anisotropic diffusion due to molecular non-sphericity is difficult to estimate. We can, however, qualitatively examine the effects of non-Brownian motion by considering the case of anomalous diffusion. In this case, the diffusion radius is given by

$$r_{\mathrm{diff}}^{anomalous}(\vec{r}, \tilde{\tau}) = \sqrt{\Gamma\ D_{\mathrm{T}}(\vec{r})\ \tilde{\tau}^{\alpha}}, \qquad (S23)$$

where $\Gamma$ is a scaling factor, and $\alpha$ characterizes the diffusion process. In the case of a typical (Brownian) diffusion $\alpha = 1$, while $\alpha > 1$ is often associated with super-diffusion processes (e.g., due to cellular transport processes).

We set $\Gamma = 6$ (corresponding to 3D Brownian diffusion) since the exact value only affects the magnitude of the diffusion constant but not the decay envelope. Fig. S8d shows the transition from Brownian diffusion to super-diffusion as we increase the exponent $\alpha$, while keeping the other values constant ($D_{\mathrm{T,bulk}} = 0.3\ \mathrm{nm}^2/\mu s$, $D_{\mathrm{R,surface}} = 0.05\frac{\mathrm{rad}^2}{\mu s}$, $z_{\mathrm{ads}} = 1.5\ \mathrm{nm}$).

The change to super-diffusion leads to a faster signal decay, while maintaining an overall shape qualitatively similar to free diffusion. This leads to a certain ambiguity in determining the diffusion regime, as the faster signal decay can be partly compensated by reducing the value of $\Gamma$ and/or

$D_{\mathrm{T,bulk}}$.

## F. Role of NV spin-lattice relaxation

The observation of a correlation signal depends on the NV ability to store the phase picked up during the first interrogation segment throughout the evolution interval $\tilde{\tau}$. A question of interest is, therefore, whether NV spin-lattice relaxation $T_{1NV}$ must be taken into account in our modeling.

While we did not specifically determine $T_{1NV}$ for the NVs we used in our correlation measurements, observations from virtually identical NVs are presented in Fig. S9. The plotted data points correspond to the difference between pump-probe and inversion-recovery protocols. Out of the 25 NVs we studied, we determine an average spin-lattice relaxation time of ~350 μs with a maximum dispersion between $T_{1NV}$ values of about 100 μs. These relaxation times are substantially longer than the time constants describing the decay of the correlation signal in the solid samples (~20 μs and ~30 μs for Samples A and B, respectively), confirming the above conclusion that NV relaxation can be ignored in these two cases. Sample C, on the other hand, deserves some special consideration: The first segment of the correlation signal (governed by molecular diffusion and decaying on a time scale of $T_{corr}^{(short)} \sim 15$ μs) is clearly insensitive to NV spin lattice relaxation, meaning that no corrections are required in our numerical estimates of the bulk diffusion constant $D_{T,bulk}$ and the adsorbed layer thickness $z_{ads}$. This is, however, not the case for the longer-lived segment of the correlation signal, which decays on a time scale $T_{corr}^{(long)} \sim 250$ μs comparable to $T_{1NV}$ (Fig. S10b). Therefore, our estimate of the rotational diffusional constant in the adsorbed layer $D_{R,ads}$ must be interpreted as a lower bound.

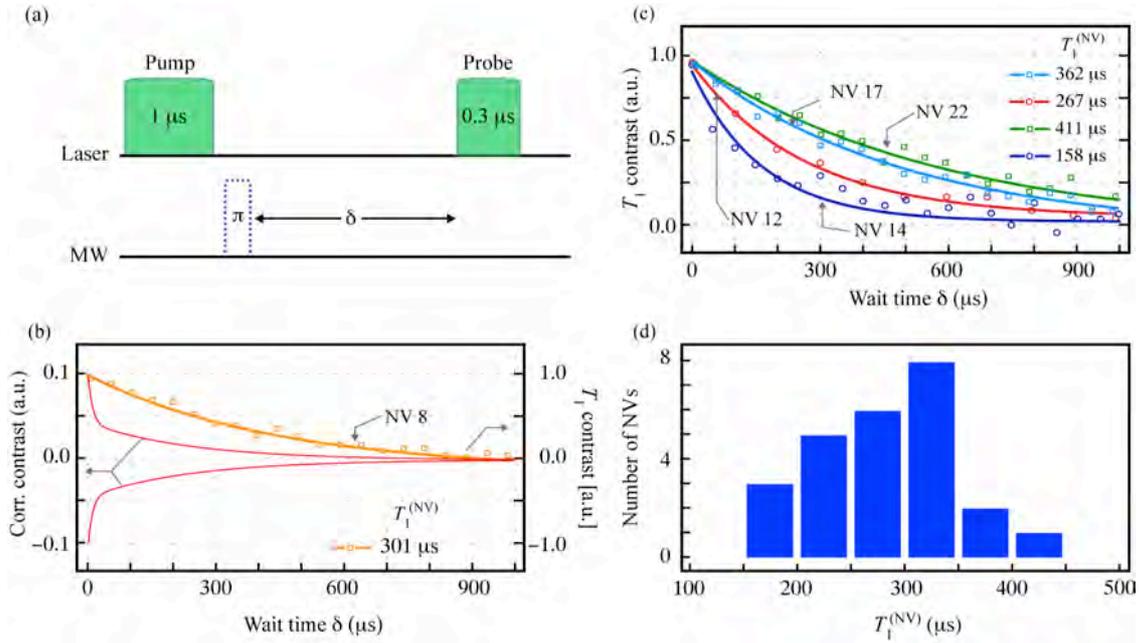

**Figure S9. Impact of NV spin-lattice relaxation on the correlation signal.** (a) We determine $T_1^{(NV)}$ by subtracting the NV signals from pump-probe and inversion recovery protocols (both differ only on the application of a π-pulse immediately after the pump laser pulse). (b) Representative $T_1^{(NV)}$ measurement (squares); the solid line is a fit to an exponential decay with time constant $T_1^{(NV)} = 301$ μs. For comparison, the red trace reproduces the (extrapolated) envelope of the correlation signal for Sample C; the long-lived section of the curve decays with a characteristic time $T_{corr}^{(long)} \sim 250$ μs. (c) Out of the 25 NVs investigated, most NVs spin-relax with a characteristic time of ~300 μs. The data sets in the plot are to be considered representative of the NV ensemble. (d) Histogram with the distribution of NV spin-lattice relaxation times.

## G. Comparison with bulk NMR data

Relevant information on the sample dynamics near the diamond surface can be gained by comparing the measured correlation signals with bulk [1]H NMR data. Fig. S10 shows the NMR proton signal for Sample A (Merckoglas) and Sample B (PDMS) upon application of a π/2-pulse at 600 MHz (corresponding to a 14.1 T magnetic field). In both cases we find that the NV-detected correlation signals (Figs. S1 and S2) are shorter-lived than their inductively detected counterparts ([1]H NMR FIDs in Figs. S10a and S10c). The difference is starker in the case of Sample B, where the nuclear spin coherence time is seen to persist beyond 0.5 ms. We interpret these differences as a manifestation of the singular molecular dynamics near the diamond surface, where motion is likely more restricted than in the bulk.

Indirect support for this idea is provided by the results of Figs. S10b and S10d where we expose the importance of molecular packing in the dynamics of both Sample A and Sample B: In the first case (Fig. S10b), we let a sample of 'as-purchased' Merckoglas dry to form a solid crust, which we then analyze via inductive NMR. This system differs from the one used in Figs. S10a only in its preparation protocol: Unlike the practice in our NV experiments—where Mercoglas is dissolved in toluene to form a 1:1 solution prior to gluing the diamond crystal to the sample holder—no solvent was added. The result is an [1]H FID substantially longer than that in Fig. S10a, the increased mobility likely originating from the reduced number of interstitial molecules after solvent evaporation. Interestingly, we observe the converse effect in Sample B where we shorten the FID duration by bringing the ratio between PDMS and the curing agent to a value lower than that used for our NV experiments. The latter leads to a more rigid form of the resulting polymer and thus a shorter-lived proton FID.

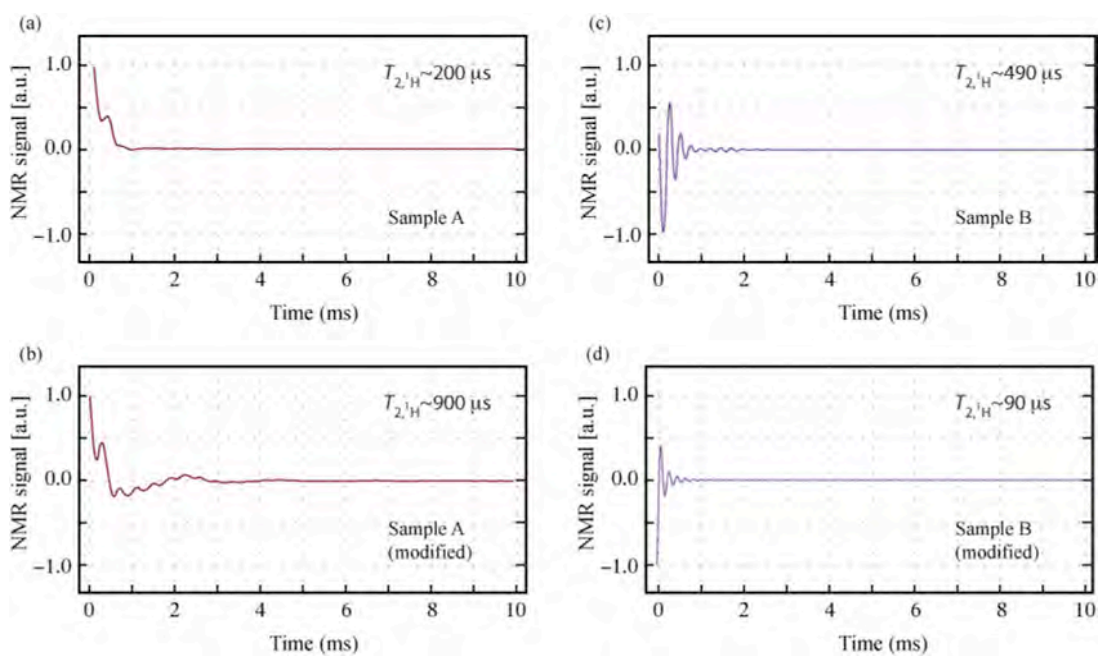

**Figure S10. High-field NMR of the solid samples.** (a) [1]H NMR free induction decay from a dried 1:1 solution of Merckoglas in toluene (as used in our NV experiments). (b) Same as in (a) but for dried Merckoglas (without added toluene). (c) [1]H NMR signal from bulk PDMS; the elastomer/curing agent ratio is identical (10:1) to that used in our NV experiments. (d) Same as in (c) but for 'rigid' PDMS (6:1 mixing ratio). All NMR experiments were carried out at 14.1 T (600 MHz proton frequency) under ambient conditions.

For completeness, Fig. S11 presents [1]H NMR data corresponding to Sample C (immersion oil), where we find a proton spin transverse relaxation time of order 25 ms, limited by field inhomogeneity (an FID from acetone is included as a reference).

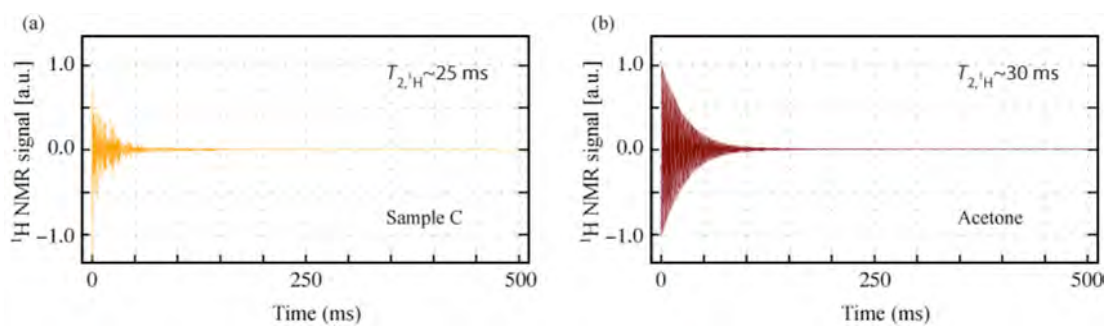

**Figure S11. $^1$H NMR signal from Sample C.** (a) $^1$H free induction decay from a sample of the immersion oil used in our NV experiments. The proton spin transverse relaxation time is approximately 25 ms, limited by field inhomogeneity. (b) FID from protons in acetone. All experiments were carried out at 14.1 T (600 MHz proton frequency) under ambient conditions.

## H. Origin of the signal

Recently it was pointed out that the XY8-N noise spectroscopy is prone to measure spurious effects of higher Larmor frequency harmonics[12]. Especially the 4th harmonic of the $^{13}$C nuclei intrinsic in the diamond lattice is spectrally located very close to the proton Larmor frequency, and can be easily misinterpreted as the signal of external protons. Our protocol is immune to this problem because it directly records the proton spin precession during $\tilde{\tau}$, thus leading to resonance spectra at the specific nuclear Larmor frequency[13]. In particular, any residual effect from carbon spins during the XY8-N segments of the correlation protocol would lead to a peak at the carbon spin resonance frequency, which can be easily separated from proton-induced contributions.